\documentclass[journal=jacsat,manuscript=article]{achemso}

\usepackage[version=3]{mhchem} 
\usepackage{amssymb,stackengine,graphicx}
\usepackage{siunitx}
\usepackage[dvipsnames]{xcolor}

\author{Vincenzo Calabrese}
\email{vincenzo.calabrese@oist.jp}
\author{Stylianos Varchanis}
\author{Simon J. Haward}
\author{Amy Q. Shen}
\email{amy.shen@oist.jp}
\affiliation[Unknown University]
{Okinawa Institute of Science and Technology, Onna-son, Okinawa 904-0495, Japan}

\title[An \textsf{achemso} demo]
  {Alignment of colloidal rods in crowded environments }

\begin{document}

\begin{abstract}
Understanding the hydrodynamic alignment of colloidal rods in polymer solutions is pivotal for manufacturing structurally ordered materials. How polymer crowding influences the flow-induced alignment of suspended colloidal rods remains unclear when rods and polymers share similar length-scales. We tackle this problem by analyzing the alignment of colloidal rods suspended in crowded polymer solutions, and comparing against the case where crowding is provided by additional colloidal rods in a pure solvent. We find that the polymer dynamics govern the onset of shear-induced alignment of colloidal rods suspended in polymer solutions, and the control parameter for the alignment of rods is the Weissenberg number, quantifying the elastic response of the polymer to an imposed flow. Moreover, we show that the increasing colloidal alignment with the shear rate follows a universal trend that is independent of the surrounding crowding environment. Our results indicate that colloidal rod alignment in polymer solutions can be predicted based on the critical shear rate at which polymer coils are deformed by the flow, aiding the synthesis and design of anisotropic materials. 
\end{abstract}

\newpage
\section{Keywords}
Cellulose nanocrystals, polymer rheology, flow-induced birefringence, rotational diffusion, colloidal rod alignment, Weissenberg number, P\'eclet number.  

\section{Introduction}
The ability to control the hydrodynamic alignment of colloidal rods is critical to produce structurally ordered soft materials that possess desirable mechanical, thermal, optical, and electrical properties.\cite{Li2021,Mittal2018,Shen2021,Xin2019,SydneyGladman2016,Sano2018,Peng2018,Hakansson2014} These anisotropic materials are promising in applications ranging from electronic sensors and soft robotics, to tissue engineering and biomedical devices. \cite{SydneyGladman2016,Pei2015,Kiriya2012,DeFrance2017} 
In material science and engineering, colloidal rods are used in combination with other polymers that impart specific functionality to the final composite material (e.g., increasing ductility and mitigating embrittlement).\cite{Kokkinis2015,Wanasekara2017,Li2021}
As such, understanding and controlling the hydrodynamic alignment of colloidal rods in polymer matrices becomes of pivotal importance in large-scale processing operations. 

Existing literature has shown that the most important control parameter for the onset of hydrodynamic alignment of rigid colloidal rods is the P\'eclet number $Pe=\dot\gamma/Dr$, a dimensionless number quantifying the relative strength between the imposed deformation rate (e.g., the shear rate, $\dot\gamma$) and the rotational diffusion coefficient of the rods ($Dr$).\citep{Mdoi_Dilute_Chapter,Lang2016,Corona2018,Lang2019,Calabrese2021} In dilute suspensions, the P\'eclet number can be defined as $Pe_0=\dot\gamma/Dr_0$, with the rotational diffusion coefficient of the rods given as
\begin{equation}
   Dr_0=\frac{3 k_b T \mathrm{ln}(l/d)}{\pi \eta_{s} l^3}~,
   \label{eqn:Dr}
\end{equation}
%
where $d$ and $l$ are the hydrodynamic  diameter and length of the colloidal rod, respectively, $k_b=1.38\times 10^{-23}$~J/K is the Boltzmann constant, $T$ is the absolute temperature, and $\eta_s$ is the solvent viscosity. For $Pe_0<1$ Brownian flocculation dominates, whilst at $Pe_0\geq 1$ convective forces are strong enough to induce alignment of the colloidal rods in the flow direction. However, this criterion is only valid under the assumption that the colloidal rods perceive the surrounding fluid as a continuum medium, i.e., the characteristic length scale of the colloidal rods, such as the radius of gyration $Rg_r=\sqrt{\frac{d^2}{2}+\frac{l^2}{12}}$, must be much larger than that associated with the suspending medium.\cite{Wyart2000,Cai2011,Koenderink2003} Colloidal rods suspended in low $Mw$ solvents such as water generally satisfy this assumption. However, in many industrial and biological processes, colloidal rods flow in crowded environments of polymers in solution where the characteristic length scale of suspended rods is similar to those of the surrounding macromolecules, e.g., the polymer radius of gyration, $Rg_p$, or the polymer mesh size, $\xi_p$ (also referred to as the correlation length).\cite{Cai2011,Alam2014,Hess2020,Gratz2019,Smith2021,Johnson1990,Lee2017a} %
In this scenario, the continuum assumption breaks down and the rods experience a local viscosity ($\eta_{local}$) that lies between the solvent viscosity and the bulk viscosity of the polymeric solution.\cite{Hess2020,Lee2017a,Alam2014,Gratz2019} In principle, by knowing $\eta_{local}$ it is possible to predict the shear rate for the onset of colloidal alignment based on the criterion of $Pe_0=1$ using $\eta_{local}$ in place of $\eta_s$ in eqn.\ref{eqn:Dr}. However, the main hurdle in predicting the alignment of colloidal rods suspended in macromolecular solutions based on the criterion $Pe_0=1$ stems from the fact that the value of $\eta_{local}$ is not known \textit{a priori}. Consequently, to date, it is challenging to predict the onset of flow-induced alignment of colloidal rods with comparable length scales to the suspending polymeric fluids. 

The flow-induced alignment of rigid elongated particles suspended in viscoelastic polymeric solutions has been studied experimentally and numerically for particles with relevant length scales larger than that associated with the suspending fluid; thus considering the suspending polymeric fluid as a continuum medium for the colloidal rods.\cite{Iso1996,Johnson1990,Hobbie2003,Gunes2008,DAvino2015,DAvino2019} In this length-scale context, theories predict a critical deformation rate above which the elastic forces of the suspending fluid cause the particle alignment to drift from the flow direction to the vorticity direction.\citep{Leal1975,Cohen1987,Iso1996,Johnson1990} However, contrasting experimental results have been reported for relatively small particles suspended in shear-thinning polymeric fluids.\citep{Johnson1990,Hobbie2003,Gunes2008} For instance, elongated hematite particles (with $l=600$~nm) in entangled polyethylene oxide solutions\cite{Gunes2008} displayed the particle alignment in the vorticity direction as expected by theory. However, shorter hematite particles (with $l=360$~nm) suspended in entangled polystyrene solutions\cite{Johnson1990} did not, casting doubts on the validity of continuum theories in conditions where colloids and polymers have similar characteristic length-scales. 

In this work, we elucidate the mechanism driving the onset of colloidal rod alignment in semi-dilute polymer solutions where polymers act as the crowding agents to tracer colloidal rods. We use cellulose nanocrystals (CNC) as rigid colloidal rods as they are widely used in synthesis of composite polymer materials with high performance and functionalities.\cite{Hasegawa2020,Eichhorn2011,Wanasekara2017} To understand the effect of crowding on the CNC alignment, we adopt two approaches: (i) increasing the CNC mass fraction, $\phi$, in an aqueous Newtonian solvent spanning from the dilute regime where interparticle interactions are negligible up to the semidilute regime where interparticle interactions are at play, providing \emph{self-crowding} of the CNC by other analogous particles; (ii) Using shear-thinning (non-Newtonian) polymer solutions as the suspending fluid, while keeping $\phi=1.0 \times 10^{-3}$ so that the CNC is in the dilute regime with negligible interparticle interactions, and the confinement acting on the CNC is provided only by the surrounding polymer chains, which we refer to as \emph{polymer-crowding}. Specifically, we use high $Mw$ neutral polymers (polyethylene oxide, PEO, and polyacrylamide), with a radius of gyration ($Rg_p$) comparable to the CNC length scale (i.e., $Rg_r$).\cite{Devanand1991,Holyst2009,Francois1979,Senses2016,Hess2020} For the polymer crowding case, we show that the onset of CNC alignment is linked with the relaxation time of the polymer solution ($\tau_p$). Specifically, we show that the Weissenberg number $Wi=\tau_p\dot\gamma$, quantifying the strength of the elastic response of the fluid to an imposed deformation rate, controls the onset of CNC alignment in polymeric media. 

\begin{figure}[ht!]
\includegraphics[width=16cm]{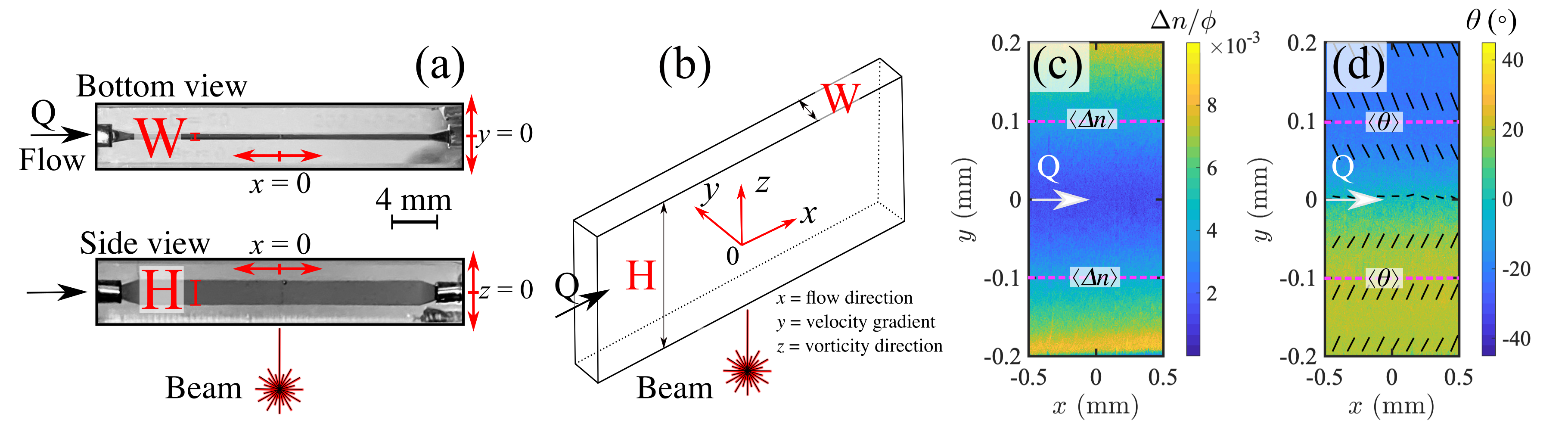}
\caption{(a) Snapshots of the microfluidic platform from bottom and side view. (b) Schematic drawing with coordinate system and relevant dimensions. (c, d) Time averaged flow-induced birefringence (FIB) across the channel width for a representative test fluid composed of CNC at $\phi=1\times10^{-3}$ suspended in water at an average velocity $U=1.6 \times 10^{-3}$~m~s$^{-1}$. In (c) the normalized  birefringence intensity ($\Delta n / \phi$) and in (d) the orientation of the slow optical axis ($\theta$),  given by the contour plot (in full resolution) and depicted by the solid segments to guide the eye. The horizontal dashed lines in (c) and (d) indicate the location where the spatially averaged birefringence, $\langle \Delta n \rangle$, and the spatially averaged angle of slow axis $\langle\theta\rangle$ are obtained for quantitative analysis.}
  \label{fgr:0}
\end{figure}
\section{Experimental section} 
\subsection{Test fluids}
The test fluids were prepared using an aqueous CNC stock suspension (CelluForce, Montreal, Canada, pH 6.3 at 5.6 wt\%). The CNC has an average length $\langle l \rangle = 260\pm180$~nm, a maximum length $l_{max}=700$ nm, an average diameter $\langle d \rangle = 4.8\pm1.8$~nm as detected from atomic force microscopy.\citep{Calabrese2021} The effective hydrodynamic diameter is computed as $d=\delta + \langle d \rangle=27.4$~nm, considering the estimated contribution of the electric double layer $\delta = 22.6$~nm in deionised water.\citep{Bertsch2017} Assuming a cylindrical shape, the number density of the CNC was calculated as $\nu=(4~\phi_{volume})/(\langle d \rangle^2 ~l~  \pi )$ where $l$ is the hydrodynamic length of the CNC obtained experimentally through $Dr_0$ (specified in the main text), and $\phi_{volume}$ is the volume fraction of the suspended CNC (calculated using a CNC density of 1560~Kg/m$^3$).\citep{Wagner2010} CNC suspensions at different mass fraction, $\phi$, were prepared by dilution of the mother CNC stock with deionised water and mixed on a laboratory roller for at least 24~h at $\sim$ 22 $^\circ$C. Where specified, the $\phi=1\times10^{-3}$ CNC suspension was prepared in a Newtonian solvent composed of a glycerol:water mixture containing 17.2 vol\% glycerol (Sigma-Aldrich 99\% with $\eta_s=$~1.7 mPa~s as measured via shear rheometry). 

Polyethylene oxide with $Mw\approx 4$~MDa, polyethylene oxide with $Mw\approx 8$~MDa, and polyacrylamide with $Mw\approx 5~$MDa, referred to as PEO4, PEO8 and PA5, respectively, were purchased from Sigma-Aldrich in powder form and solubilized in deionised water on a laboratory roller for at least 48~h at $\sim22$~$^\circ$C (stock solution). Polymer solutions at different concentrations ($c$ in mg/mL) containing a constant amount of CNC were prepared by diluting the polymer stock solution with deionised water followed by dilution of the CNC stock suspension to a final $\phi = 1 \times 10^{-3}$, followed by mixing on a laboratory roller for 24~h at $\sim22$~$^\circ$C. Polymer solutions without the CNC were prepared by following the same procedure described above. The polymer concentration where polymers begin to overlap was estimated as $c^*=Mw/((4\pi /3) Rg_p^3 N_A)$, where $N_A$ is the Avogadro's number. The radius of gyration for the PEO4 and PEO8 was 135 and 202~nm, respectively, estimated as $Rg_p=0.02 Mw^{0.58}$.\citep{Devanand1991,Holyst2009} For PA5 $Rg_p=7.5\times10^{-3} Mw^{0.64}=154$~nm.\citep{Francois1979} For the PEO8, the polymer concentrations tested spanned between the dilute ($c/c^*< 1$) and semidilute unentangled regime, where $c/c^*> 1$ and the tube diameter $t_d=b_k \phi_p^{-0.75}>Rg_p$, where $\phi_p$ is the polymer volume fraction and $b_k\approx 3.7$~nm is the experimentally determined tube diameter in a PEO melt.\cite{Senses2016,Hess2020,rubinstein2003polymer}  For PEO8 with $Rg_p = 202$~nm, the highest polymer concentration tested is $c/c^*=10.5$ which yields $t_d\approx 250$~nm, thus in the semidilute unentangled regime ($c/c^*>1$ and $t_d>Rg_p$). 

\subsection{Methods}
To assess the CNC alignment in crowded environments, we measure the flow-induced birefringence (FIB) as a function of the shear rate in a straight microfluidic channel etched in fused silica (Fig.~\ref{fgr:0}(a,b)). The microfluidic channel has a rectangular cross section with height $H=2$~mm along the $z$-axis, corresponding to the optical path, and width $W=0.4$~mm, along the $y$-axis, thus providing an approximation to a two-dimensional (2D) Poiseuille flow.\citep{Haward2018,Calabrese2021} The flow in the microfluidic channel is driven along the channel length (\textit{x}-axis) by a syringe pump (Cetoni Nemesys) and Hamilton Gastight syringes infusing and withdrawing at an equal and opposite volumetric flow rate ($Q$) from the inlet and outlet, respectively. The average flow velocity in the channel is $U = Q(WH)^{-1}$.

\subsubsection{Flow-induced birefringence (FIB)}
Time-averaged FIB measurements were performed using an Exicor MicroImager (Hinds Instruments, Inc., OR) using a 5$\times$ objective at room temperature ($\sim22$~$^\circ$C). The channel is illuminated through the $z$-axis (vorticity direction) using a monochromatic beam of wavelength $\lambda=450$~nm  or $\lambda=630$~nm (see Fig.~\ref{fgr:0}(a,b)). The retardance, $R$ in nm, and the orientation of the slow optical axis (extraordinary ray), $\theta$, were obtained from 7 images acquired at 1 s interval and time-averaged. The spatially-resolved  (spatial resolution of $\Delta n $ and $\theta$ is $\approx 2$ \SI{}{\micro\meter}/pixel), and time-averaged, birefringence ($\Delta n = R/H$) and the orientation of the slow optical axis, $\theta$, are obtained in the flow-velocity gradient plane ($x$-$y$ plane) as shown in Fig.~\ref{fgr:0}(c) and (d), respectively, for a representative test fluid of dilute CNC in water. For each flow rate, the birefringence $\Delta n$ and the absolute value of the orientation of the slow optical axis, $|\theta|$, are spatially averaged along 1~mm of the $x$-axis at $y = \pm 0.1~$mm (see dashed lines in Fig.~\ref{fgr:0}(c, d)) and referred to as $\langle \Delta n \rangle$ and $\langle \theta \rangle $, respectively. The background value of $\Delta n$ acquired at rest was subtracted for all the analysis presented and quantitative analysis of $\theta$ is restricted to $\Delta n > 3 \times 10^{-6}$. The orientation angle $\langle \theta \rangle $ probes the CNC orientation with respect to the flow direction whilst the birefringence intensity, $\langle \Delta n \rangle$, probes the extent of anisotropy in the system. For fully isotropically oriented particles $\langle \Delta n \rangle = 0$ with colloidal alignment occurring for $\langle \theta \rangle < 45^\circ$ and $\langle \Delta n \rangle > 0$. \cite{Vermant2001,Reddy2018} The error related to the spatially averaged $\langle \Delta n \rangle$ and $\langle \theta \rangle$ is the standard deviation from the averaging process.

\subsubsection{Rheology and flow simulations}
Shear rheometry of the test fluids was performed using a strain-controlled ARES-G2 rotational rheometer (TA Instrument Inc.) equipped with a stainless steel cone and plate geometry (50 mm diameter and 1$^\circ$cone angle). The test fluids were covered with a solvent trap and measured at 25$\pm$0.1 $^\circ$C (controlled by an advanced Peltier system, TA Instruments). The shear viscosity data were fitted to the Carreau-Yasuda (CY) generalized Newtonian model: 
\begin{equation}
   \eta=\eta_\infty +\frac{\eta_0-\eta_\infty}{[1+(\dot\gamma / \dot\gamma^*)^a]^{(1-n)/a}}~,
   \label{eqn:Carreau_Yasuda}
\end{equation}
where $\eta_0$ is the zero-shear-rate viscosity, $\eta_{\infty}$ is the infinite-shear-rate viscosity, $\dot\gamma^*$ is the characteristic shear rate for the onset of shear thinning, $n$ is the power law exponent and $a$ is a dimensionless fitting parameter that controls the transition to the shear-thinning region. The velocity field along the channel width ($W$, $y$-axis) and the value of shear rate at $y=|0.1|$~mm, (the channel location where $\langle \Delta n \rangle$ and $\langle \theta \rangle$ are obtained) were computed using numerical simulation. The simulations were performed assuming steady, 1-dimensional, fully developed flow in a planar channel with width $W$, under the imposition of an average velocity $U$. The $x$-component of momentum equation and the CY constitutive equation were discretized and solved using an in-house Finite Element solver \cite{varchanis2019new}. In all simulations, 200 linear elements were used across the width of the channel. For each average velocity used during the FIB experiment, $\langle \Delta n \rangle$ and $\langle \theta \rangle$ are compared with the effective value of $|\dot\gamma|$ obtained from the simulation at $y=|0.1|$~mm. The channel location of $y=\pm 0.1$~mm is chosen to be the midpoint between the side walls ($y= \pm 0.2$~mm) and the centerline ($y=0$~mm) so to provide relatively high values of shear rate while avoiding undesired wall effects in the FIB experiment. For very weakly shear-thinning fluids, for which the CY model could not be fitted to the rheological data, the shear rate was computed as for a Newtonian fluid.

\section{Results and discussion}

\begin{figure}[ht!]
\includegraphics[width=16.7cm]{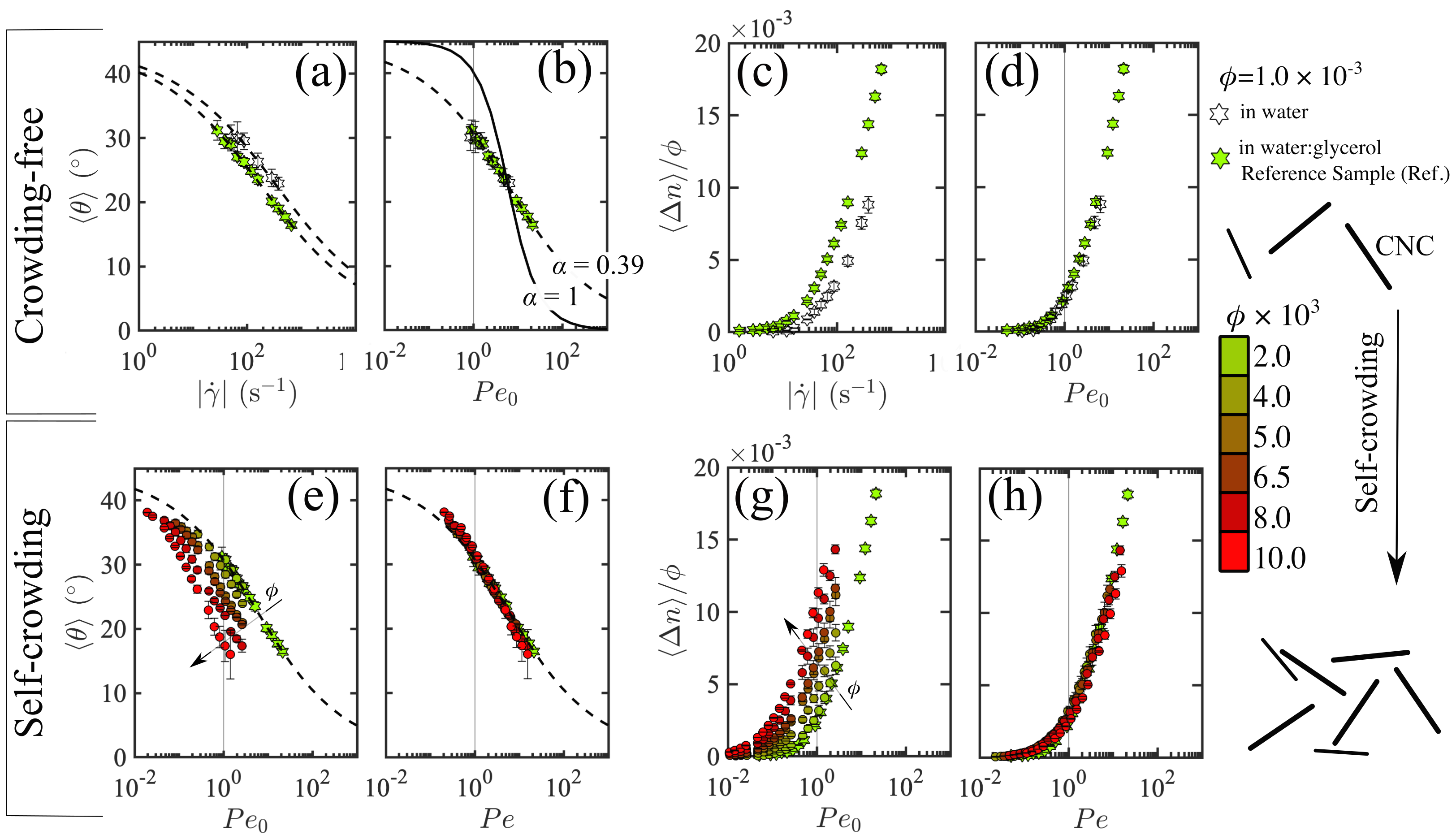}
\caption{Spatially averaged orientation angle $\langle \theta \rangle$ (a, b, e, f) and normalized birefringence $\langle \Delta n \rangle/\phi$ (c, d, g, h) for CNC suspensions in the crowding-free (a-d) and self-crowding (e-h) regime. Open and green filled stars represent CNC in water and in water:glycerol mixture (17.2 vol\% glycerol), respectively (crowding-free). For the samples prepared in water the CNC rotational diffusion coefficient in the crowding-free regime is $Dr_0=59$~s$^{-1}$ whilst $Dr_0=31$~s$^{-1}$ for the sample prepared in the water:glycerol mixture labeled by the green star symbol and used as a reference sample (Ref.). For the crowding-free regime, $\langle \theta\rangle$ (a) and $\langle \Delta n \rangle /\phi$ (c) as a function of $|\dot\gamma|$ and scaled as $Pe_0=|\dot\gamma|/Dr_0$ in (b) and (d), respectively. For the self-crowding regime, $\langle \theta\rangle$ (e) and $\langle \Delta n \rangle /\phi$ (g) as a function of $Pe_0$ and scaled as $Pe=|\dot\gamma|/Dr$ in (f) and (h), respectively. The dashed lines in (a, b, e, f) are the fittings to eqn.~\ref{eqn:Orientation Angle} with $\alpha=0.39$. In (b) eqn. \ref{eqn:Orientation Angle} is also plotted using $\alpha=1$ (solid line). Vertical lines at $Pe_0=1$ and $Pe=1$ are drawn as a reference. Error bars indicate the standard deviation of the measurement.}
  \label{fgr:1}
\end{figure}

\subsubsection{Crowding-free}
We begin with the evaluation of the FIB of dilute CNC at $\phi=1.0 \times 10^{-3}$ suspended in an aqueous Newtonian media, either in water ($\eta_s =0.9$~mPa~s) or water:glycerol mixture (17.2 vol\% glycerol, $\eta_s= 1.7$~mPa~s), shown in Fig.~\ref{fgr:1}(a-d). At this relatively low CNC concentration, interparticle interactions are negligible so that each CNC can be considered as isolated and crowding-free.\cite{Bertsch2019} Fig.~\ref{fgr:1}(a) and (c) present the $\langle \theta \rangle $ and $\langle \Delta n \rangle/\phi$ as a function of $|\dot\gamma|$ for the CNC in the crowding-free regime respectively.  For both media, the orientation angle, $\langle \theta \rangle $, displays a gradual decrease with $|\dot\gamma|$ (Fig.~\ref{fgr:1}(a)) and, for a given value of $|\dot\gamma|$, the greater solvent viscosity of the water:glycerol mixture favours the CNC to align with a smaller value of $\langle \theta \rangle $ compared to water as the solvent. Moreover, the greater viscosity of the water:glycerol media triggers the onset of CNC alignment at a lower shear rate than in water, as indicated by the onset of birefringence occurring at lower values of $|\dot\gamma|$ (Fig.~\ref{fgr:1}(c)).

The effective rotational diffusion coefficient $Dr$ of the CNC is obtained experimentally based on
\begin{equation}
   \langle \theta \rangle=\frac{\pi}{4} - \frac{1}{2} \mathrm{arctan}\bigg[\bigg(\frac{ |\dot\gamma|}{6 Dr}\bigg)^\alpha \bigg]~, 
   \label{eqn:Orientation Angle}
\end{equation}
where $0 < \alpha \leq 1$ is a stretching exponent that accounts for particle polydispersity.\cite{Reddy2018,Lindsey1980,Thurn1984,Decruppe2003} For monodisperse particles $\alpha =1$ whilst for polydisperse particles $\alpha<1$ (see solid and dashed line in Fig.~\ref{fgr:1}(b)). Fitting the data in Fig.~\ref{fgr:1}(a) with eqn.~\ref{eqn:Orientation Angle} yields $\alpha=0.39$ for both Newtonian solvents, while $Dr_0=59$~ s$^{-1}$ and $Dr_0=31$~s$^{-1}$ for the water and water:glycerol solvents respectively. The subscript ``0'' to $Dr$ indicates the crowding-free regime (i.e., dilute CNC in a continuum medium) and is used to distinguish it from the case where $Dr$ is affected by the surrounding crowding agent. 
By solving eqn.\ref{eqn:Dr} for $l$, using an effective CNC diameter $d = 27.4$~nm accounting for the contribution from the electric double layer,\citep{Bertsch2017} and $T=22~^{\circ}$C, we obtain $l=610$~nm, in accordance with the longest CNC population detected via microscopy in our previous work.\cite{Calabrese2021}
Our experimental data collapse onto master curves when we plot $\langle \theta \rangle $ and $\langle \Delta n \rangle /\phi$  as a function of the respective P\'eclet number in the crowding-free regime, $Pe_0=|\dot\gamma|/Dr_0$, and, the birefringence signal increases sharply at $Pe_0=1$, indicating that the $Pe_0$ scaling is correct (Fig. \ref{fgr:1}(b,~d)). It is expected that the $Pe_0$ scaling yields master plots of $\langle \theta \rangle $ and $\langle \Delta n \rangle /\phi$ only for the systems where the CNC is not experiencing any confinement.   
It is important to note that, $\langle \Delta n \rangle$ and $\langle \theta \rangle$ are complementary parameters to quantify particle alignment during flow. The magnitude of birefringence $\langle \Delta n\rangle$ is linearly related with the number of aligned particles in a given illuminated volume, thus presented in a normalized form as $\langle \Delta n \rangle /\phi$.  Contrarily, for a uniform particle alignment, $\langle \theta \rangle$ captures a geometrical property of the system that is independent  from the number of aligned particles in a given illuminated volume.\cite{Vermant2001}

\subsubsection{Self-crowding}
Following the same approach as for the crowding-free case, we investigate the CNC alignment upon increasing CNC mass fraction $\phi$, so that analogous particles restrain the motion of each other; a regime that we refer to as self-crowding (Fig.~\ref{fgr:1}(e,~h)). As a reference sample we use the crowding-free suspension in water:glycerol mixture (see green star symbol). Increasing the CNC concentration, the values of $\langle \theta \rangle$ decrease with increasing $\phi$ (Fig.~\ref{fgr:1}(e)). Analogously, the birefringence onset occurs at smaller values of $Pe_0$, corresponding to lower values of $|\dot\gamma|$ (Fig.~\ref{fgr:1}(g)). This behaviour can be explained by the restrained rotational motion of the CNC above the overlap concentration due to particle confinement, leading to a decreasing $Dr$ with increasing $\phi$. For each sample, $Dr$ can be obtained by fitting the curves in Fig.~\ref{fgr:1}(e) with eqn.~\ref{eqn:Orientation Angle} using a fixed value of $\alpha =0.39$ as established in the crowding-free regime. By plotting $\langle \theta \rangle$ and $\langle \Delta n \rangle /\phi$ as a function of the P\'eclet number $Pe=|\dot\gamma|/Dr$, the curves collapse onto single master curves (Fig.~\ref{fgr:1}(f, h)) and eqn.~\ref{eqn:Orientation Angle} leads to the dashed line in Fig.~\ref{fgr:1}(f). We note that for $\langle \theta \rangle$, the master curve is set by only considering $Pe$ as scaling factor. Contrarily, for the birefringence $\langle \Delta n \rangle$, the CNC mass fraction $\phi$ needs to be considered, indeed only scaling the birefringence as $\langle \Delta n \rangle /\phi$ enables to collapse the curves onto a master curve. 

\begin{figure}[ht!]
\includegraphics[width=16cm]{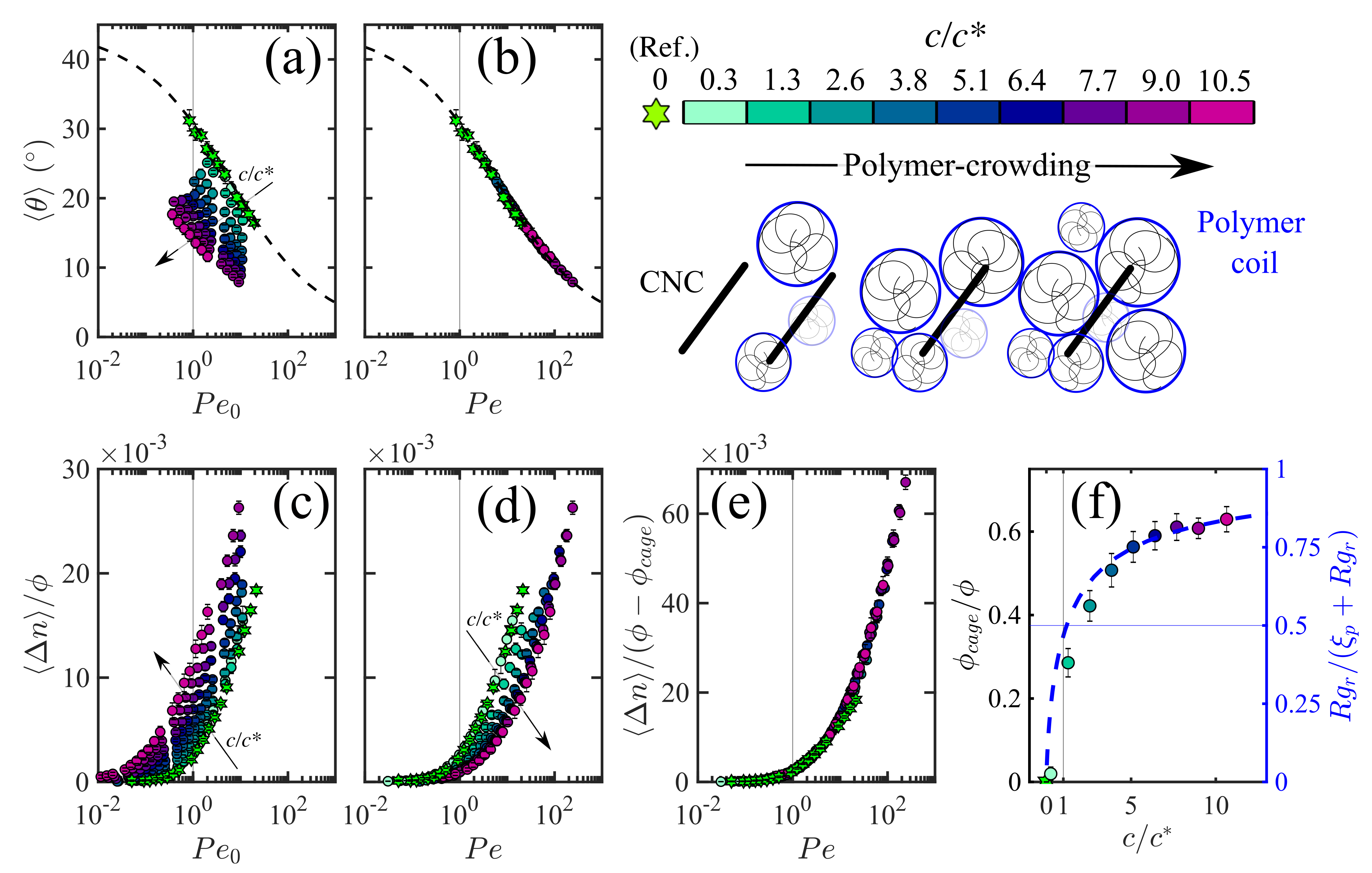}
\caption{Spatially averaged birefringence $\langle \Delta n \rangle$ and orientation angle $\langle \theta \rangle$ for PEO8 solutions, at different $c/c^*$ , seeded with CNC at $\phi= 1.0 \times 10^{-3}$. The PEO8 solutions are prepared in water, thus the rotational diffusion coefficient in the crowding-free regime is $Dr_0=59$~s$^{-1}$, whilst $Dr_0=31$~s$^{-1}$ for the reference water:glycerol mixture, labeled by the green star symbol (Ref.). Orientation angle $\langle \theta \rangle$ (a) and, normalized birefringence $\langle \Delta n \rangle /\phi$ (c) as a function of $Pe_0=|\dot\gamma|/Dr_0$ and scaled as $Pe=|\dot\gamma|/Dr$ in (b) and (d), respectively. In (e) the birefringence $\langle \Delta n \rangle$ is scaled as $\langle \Delta n \rangle/ (\phi - \phi_{cage})$ where  $\phi - \phi_{cage}=\phi_{eff}$ and plotted as a function of $Pe$. (f) The $\phi_{cage}/\phi$ (circles) and $Rg_r/(\xi_p+Rg_r)$ (dashed blue line) as a function of polymer concentration ($c/c^*$). $Rg_r$ is the radius of gyration of the CNC and $\xi_p$ is polymer mesh size (eqn.~\ref{eqn:mesh_polymer}). The dashed line in (a) is an instance of the fittings to eqn.~\ref{eqn:Orientation Angle} with $\alpha=0.39$. In (b) the curve described by eqn.~\ref{eqn:Orientation Angle} collapse on to a master curve. Vertical lines at $Pe_0=1$ and $Pe=1$ are drawn as a reference. Error bars from panel (a-e) indicate the standard deviation of the measurement, whilst in (f) error bars indicate uncertainty associated with $\phi_{cage}$ calculation.  }
  \label{fgr:2} 
\end{figure}

\subsubsection{Polymer-crowding}

For the case of polymer-crowding we follow $\langle \theta \rangle$ and $\langle \Delta n \rangle$ arising from the alignment of diluted CNC ($\phi = 1.0 \times 10^{-3}$) in polymer solutions exposed to a shearing flow. We consider polymers with different $Mw$ and concentrations to achieve viscoelastic polymer solutions with viscosity up to $400\times$ the viscosity of water, and relaxation times $0.04<\tau_p<0.7$~s, to provide a large span of crowding environments to the CNC. We use aqueous solutions of polyethylene oxide with $Mw \approx 4$~MDa and $Mw \approx 8$~MDa, referred to as PEO4 and PEO8, respectively, and polyacrylamide with $Mw \approx 5.5$~MDa, referred to as PA5.
In the absence of CNC, these polymer solutions do not display any significant birefringence, ensuring that $\langle \Delta n \rangle$ and $\langle \theta \rangle$ signals measured for the CNC dispersions arise exclusively due to the CNC alignment. Here we focus on PEO8 solutions; analysis for the PEO4 and PA5 solutions are given in the ESI~Section 2. The confinement imposed on the CNC is tuned by the polymer concentration ($c$), expressed in normalized form as $c/c^*$,  where $c^*$ is the polymer overlap concentration (Fig.~\ref{fgr:2}). For the PEO8, $c^*=0.39$~mg/mL. The PEO8 concentration in the polymer crowded CNC dispersions is varied between the dilute regime, $c/c^*<1$, and the semidilute unentangled regime, where $c/c^*> 1$ and the tube diameter ($t_d$) is greater than the PEO8 radius of gyration, $Rg_p= 202$~nm.\cite{rubinstein2003polymer} Upon increasing the PEO8 concentration, $c/c^*$, the $\langle \theta \rangle$ decreases for a given shear rate, thus $Pe_0=|\dot\gamma|/Dr_0$ (see Fig.~\ref{fgr:2}(a)), and the onset of birefringence shifts towards lower values of $Pe_0$ (Fig.~\ref{fgr:2}(c)). Interestingly, by plotting $\langle \theta \rangle$ as a function of $Pe =|\dot\gamma|/Dr$, using the values of $Dr$ obtained from the fitting of eqn.~\ref{eqn:Orientation Angle}, the curves collapse onto a master curve (Fig.~\ref{fgr:2}(b)), revealing that the alignment of the CNC occurs in a similar manner for different PEO8 concentrations. 
Elliptical hematite particles (with $l=600$~nm and cross-section of $130$~nm) in entangled PEO solutions have been reported to first orient along the flow direction and then evolve to orientations in the vorticity direction.\cite{Gunes2008} However, within our experimental window, we could only observe $\langle \Delta n \rangle / \phi$ increasing as a function of $Pe$, without the birefringence drop associated with the particle alignment drifting from the flow direction to the vorticity direction.\cite{Gunes2008,Hobbie2003,Johnson1990} Indeed, perfect alignment along the vorticity direction ($z$-axis) for particles with circular cross-sections would lead to isotropic projections in the flow-velocity gradient plane ($x$-$y$ plane) with $\langle \Delta n \rangle=0$. Similarly to our results, Johnson et al.\cite{Johnson1990} only observed the orientation in the flow direction for elliptical hematite particles (with $l=360$~nm and cross section of $100$~nm) in entangled polystyrene solutions. 

Different from the self-crowding cases, scaling $\langle \Delta n \rangle /\phi$ by $Pe$ is insufficient to collapse the data onto a single master curve (Fig.~\ref{fgr:2}(d)). We re-analyze the data based on the assumption that the failure to collapse might be caused by a small amount of caged CNC ($\phi_{cage}$) within the PEO8 matrix that remain in an isotropic state, which do not contribute to $\langle \Delta n \rangle$ intensity. 
Indeed, scaling the $\langle \Delta n \rangle$ by an effective CNC mass fraction $\phi_{eff}=\phi - \phi_{cage}$ enables the data to collapse onto a master curve (Fig.~\ref{fgr:2}(e)). For each $c/c^*$, $\phi_{eff}$ is obtained by minimizing the sum of squared residuals between the reference water:glycerol curve and the polymer-containing samples (detailed procedure is given in the ESI Section 3). The $\phi_{cage}/ \phi$ increases with $c/c^*$ and at $c/c^*\approx  5$, it seems to approach a plateau value $\phi_{cage}/\phi \approx 0.65$ (Fig.~\ref{fgr:2}(f)). To corroborate the hypothesis that a portion of CNC is entrapped within the PEO8 network without contributing to the birefringence intensity, it is instructive to understand the evolution of $\phi_{cage}$ from a topological perspective. As such, we compare the radius of gyration of the CNC, $Rg_r=\sqrt{\frac{d^2} {2} + \frac{l^2} {12}}=177$~nm, using $l=610$~nm as previously determined from $Dr_0$, and $d=27.4$~nm, with the mesh size of the PEO8 \cite{Holyst2009} 
\begin{equation}
\xi_p=Rg_p(c/c^*)^{-0.75}~,
\label{eqn:mesh_polymer}
\end{equation}
as a function of $c/c^*$, where $Rg_p=202$~nm is the radius of gyration of PEO8. Specifically, we use the ratio $Rg_r/(\xi_p+Rg_r)$ to yield values between 0 ($\xi_p\rightarrow \infty$) and 1 ($\xi_p\rightarrow0$). Both $\phi_{cage}/\phi$ and the theoretical curve $Rg_r/(\xi_p+Rg_r)$, based on eqn.~\ref{eqn:mesh_polymer}, display a similar trend as a function of $c/c^*$, suggesting that $\phi_{cage}$ is linked to the topological confinement exerted by the polymer mesh size $\xi_p$. Thus, at $c/c^*< 5$, $\xi_p$ is a strong function of the polymer concentration, and $\phi_{cage}/\phi$ increases likewise. Contrarily, at $c/c^*> 5$,  $\xi_p$ becomes a weaker function of $c/c^*$, accordingly, $\phi_{cage}$ starts to plateau. 

Overall, our observations imply that the PEO8 provides a two-way topological hindrance to the CNC rotation, where a fraction of the CNC, $\phi_{eff}$, perceives the confinement but is able to align at $Pe>1$, whilst the other remaining fraction, $\phi_{cage}$, is unable to align in the range of investigated $Pe$. Contrarily, for the case of self-crowding, all the CNC contribute to the birefringence signals, i.e., $\phi=\phi_{eff}$. This significant difference between self- and polymer-crowding is associated with the network formed from flexible PEO8 polymer chaining versus the network composed of rigid rod-like CNC. In the following sections we examine the dependence of $\phi_{eff}$ on relevant length- and time-scales.
\begin{figure}[ht!]
\includegraphics[width=10cm]{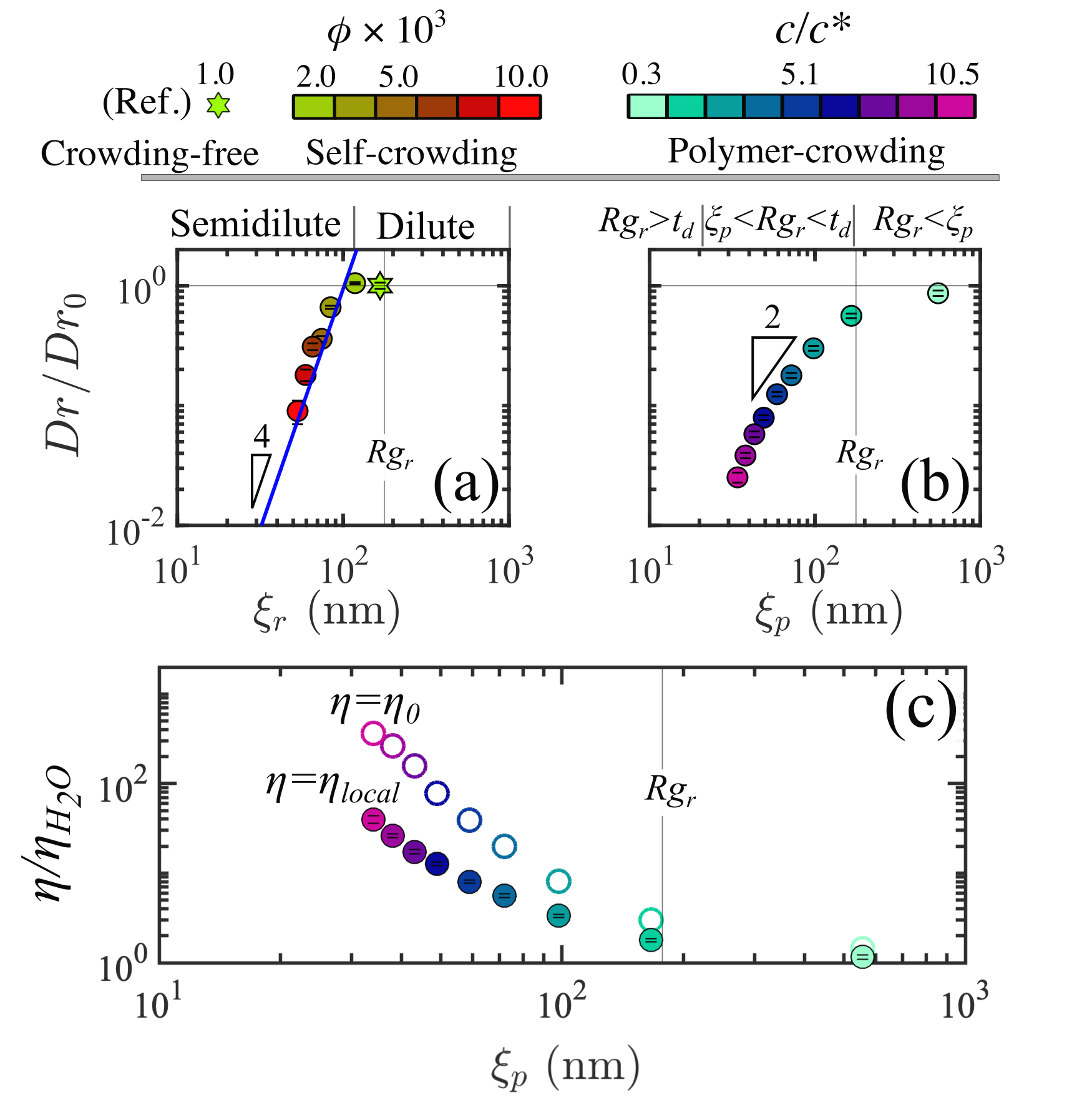}
\caption{(a,~b) Rotational diffusion coefficient of CNC, $Dr$, obtained from FIB (eqn.\ref{eqn:Orientation Angle}), normalized by the respective rotational diffusion coefficient in the crowding-free regime, $Dr_0$, as a function of the mesh size of the crowding agent $\xi$. (a) The case of self-crowding, given by an increasing CNC concentration with the corresponding mesh size $\xi_r$ (eqn.~\ref{eqn:mesh_rod}). (b) The case of polymer-crowding, given by an increasing PEO8 concentration with the corresponding mesh size $\xi_p$ (eqn.~\ref{eqn:mesh_polymer}).  Relevant regimes for the (a) self- and (b) polymer-crowding  are annexed above the respective panels. Line in (a) is the prediction from eqn.~\ref{eqn:Doi_Mesh}. In (b), the triangle indicates the scaling prediction from Cai et al.\citep{Cai2011} where $Dr\propto \xi_r^2 $ in the regime $\xi_p<Rg_r<t_d$. (c) Comparison between the normalized zero shear viscosity $\eta_0 / \eta_{H_2O}$ measured by bulk shear rheology (empty symbols), and the viscosity experienced by the CNC, $\eta_{local}/ \eta_{H_2O}$ (filled symbols), obtained by eqn.~\ref{eqn:Diffusion-viscosity}. Error bars in (a,~b) indicate uncertainty associated to $Dr$ obtained from the fitting procedure of eqn.~\ref{eqn:Orientation Angle}. 
}
  \label{fgr:3} 
\end{figure}
\subsubsection{Length-scale dependence}

The rotational diffusion coefficient $Dr$ obtained from eqn.~\ref{eqn:Orientation Angle} captures the impact of the crowding environment on the CNC. To compare the $Dr$ for the case of self- and polymer-crowding, we plot $Dr$ against the statistical mesh size $\xi$ of each crowding agent. The mesh size of the polymer matrix is given by $\xi_p$ (eqn.~\ref{eqn:mesh_polymer}), whilst for the CNC suspension it is estimated as 
\begin{equation}
\xi_{r}=(l\nu)^{-0.5}~,
\label{eqn:mesh_rod}
\end{equation}
where $\nu$ is the number density of the CNC (number of CNC per unit volume).\citep{DeGennes1976,Lang2019} For both cases $Dr$ decreases progressively with the decreasing $\xi$, indicating that the rods progressively sense the local confinement with the decreasing $\xi_r$ or $\xi_p$. However, the CNC follows a much sharper decrease in $Dr$ as a function of $\xi$ for the case of self-crowding than that of the polymer-crowding, indicating that the mesh size $\xi$ alone is not able to fully capture the dependence of $Dr$ from the crowding agent. The $Dr$ trend for the self-crowding case shown in Fig.~\ref{fgr:3}(a) meets the expectation from rigid-rod theory where in the dilute regime $Dr$ is concentration independent and $Dr/Dr_0 = 1$. Whilst in the semidilute (self-crowding) regime, the CNC motion is constrained by the surrounding rods and $Dr$ becomes concentration dependent.\cite{Mdoi_Semidilute_Chapter} The dependence of $Dr$ with the rod concentration has been described by the tube model in the framework of the Doi-Edwards theory, assuming that particles are rigid and monodisperse rods in the semidilute regime, as
\begin{equation}
Dr/Dr_0 = \beta (\nu l^3)^{-2}~, 
\label{eqn:Doi_Tube}
\end{equation}%
where $\beta$ is a length-independent prefactor.\cite{Mdoi_Semidilute_Chapter,Doi1978Part2,LangStiffness} Recently, Lang et al.\cite{Lang2019} experimentally validated for monodisperse colloidal rods that $\beta= 1.3 \times 10^3$ as previously found from computer simulation.\cite{teraoka1985molecular} Rearranging eqn.~\ref{eqn:Doi_Tube} with eqn.~\ref{eqn:mesh_rod}, we obtain 
 \begin{equation}
Dr/Dr_0 = \beta \xi_r^4 l^{-4}. 
\label{eqn:Doi_Mesh}
\end{equation}
%
Although polydisperisity is not accounted for in the Doi-Edwards theory, it is interesting to note that the $\xi_r^4$ scaling captures the $Dr/Dr_0$ trend for the self-crowding case (Fig.~\ref{fgr:3}(a)).  Moreover, using $\beta= 1.3 \times 10^3$ and $l=610$~nm in eqn.~\ref{eqn:Doi_Mesh}, it is possible to quantitatively capture the increasing $Dr/Dr_0$ with $\xi_r$ in the semidilute (self-crowding) regime (see line in Fig.~\ref{fgr:3}(a)). 

\begin{figure}[ht!]
\includegraphics[width=14cm]{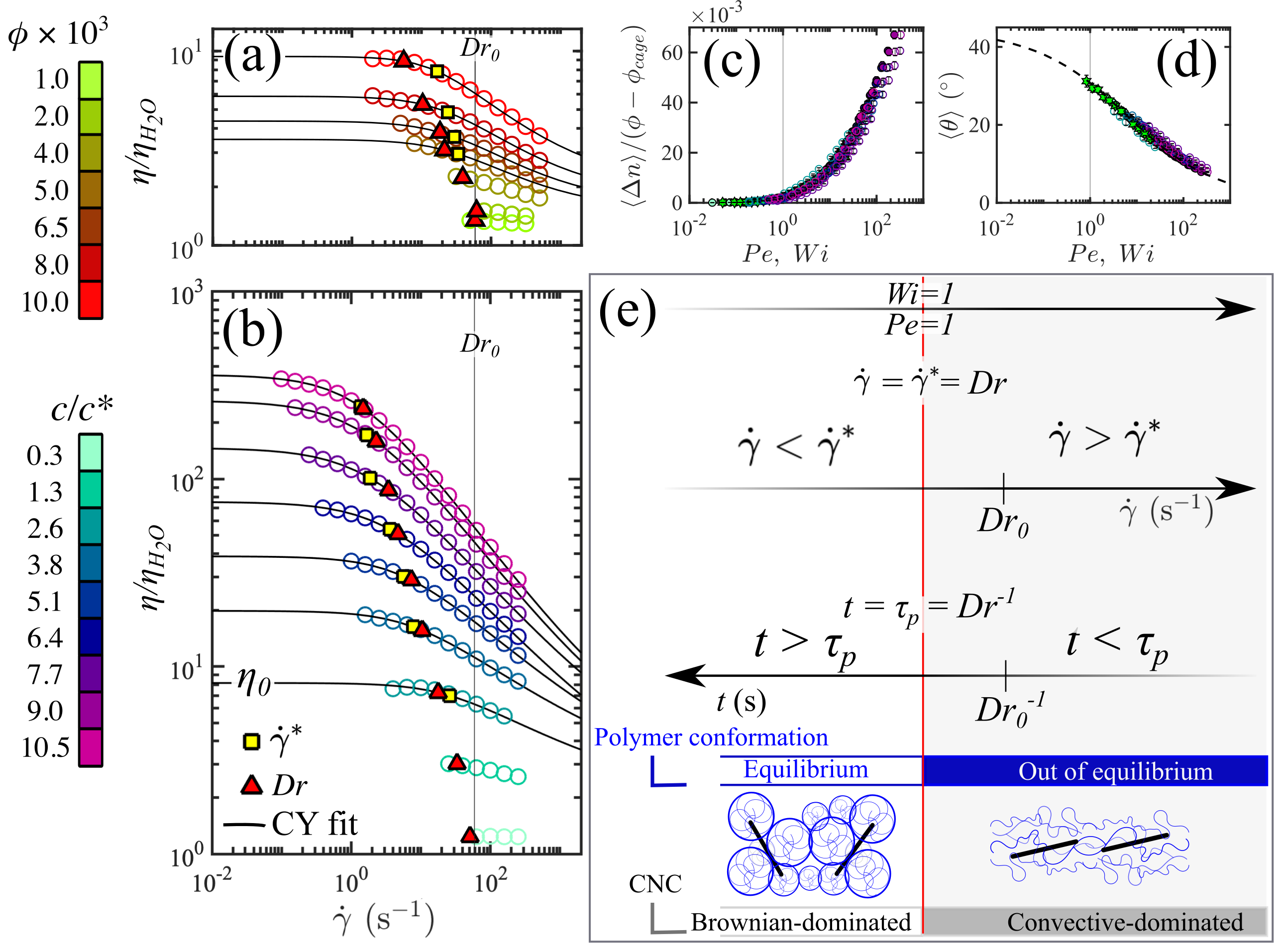}
\caption{Steady shear viscosity measurements for (a) CNC suspended in water at different values of $\phi$, and (b) PEO8 solutions at different values of $c/c^*$ (without CNC) presented as $\eta/\eta_{H_2O}$ \textit{vs.} $\dot\gamma$. Solid lines in (a, b) describe the CY model (eqn.~\ref{eqn:Carreau_Yasuda}). The onset of shear thinning, $\dot\gamma^*$, obtained from the CY, and the value of $Dr$ obtained from FIB measurements are plotted on the respective viscosity values. (c,~d) Comparison between the $Pe$ (filled symbols) and $Wi$ (empty symbols) scaling for the normalized birefringence, $\langle \Delta n\rangle /(\phi-\phi_{cage})$ , and orientation angle, $\langle \theta \rangle$, for the CNC in PEO8 solutions at different values of $c/c^*$. (e) Schematic of relevant time-scales involved in the alignment of colloidal rods in polymer-crowding cases. The schematic is valid for polymers with a relaxation time $\tau_p>Dr_0^{-1}$.  }
  \label{fgr:4} 
\end{figure}

For the polymer-crowding case, we take inspiration from prior work on the translational and rotational diffusions of nanorods in PEO solutions.\cite{Cai2011,Alam2014} 
Specifically, we adopt the scaling law developed by Cai et al.\cite{Cai2011,Alam2014}, by  considering $Rg_r$ as the characteristic dimension of CNC. The majority of the data fall in the ``intermediate regime'' where $\xi_p<Rg_r<t_d$, annotated above Fig.~\ref{fgr:3}(b). In this regime, the scaling theory predicts $Dr \propto \xi_p^2$, which captures our trend reasonably well.\cite{Cai2011} 
We have also used other polymeric solutions (PEO4 and PA5) to gain further understanding of the polymer-crowding cases. We confirm that the mesh size alone is unable to fully describe the dependence of $Dr$ from the crowding agent, thus, the curves of $Dr/Dr_0$ vs.\ $\xi_p$ do not collapse onto a master curve (see ESI, Fig. S3). 
Since CNC in the PEO8 solutions are in the dilute regime ($\phi =1.0 \times 10^{-3}$),  it is instructive to evaluate the local viscosity experienced by the particles, $\eta_{local}$, as \citep{Wisniewska2017,Gratz2019,Choi2015,Hess2020}
\begin{equation}
Dr/Dr_0=\eta_{H_2O}/\eta_{local}~.
\label{eqn:Diffusion-viscosity}
\end{equation}
We use $Dr_0=59$~s$^{-1}$ determined in water with $\eta_{H_2O}$ being the water viscosity and $Dr$ obtained experimentally from eqn.~\ref{eqn:Orientation Angle} (plotted in Fig.~\ref{fgr:3}(a)), to retrieve $\eta_{local}$. In Fig.~\ref{fgr:3}(c), the zero-shear-rate viscosity $\eta_0$ obtained from bulk rheology (details in Fig.~\ref{fgr:4}(b)) is compared to $\eta_{local}$, displaying $\eta_{local} < \eta_0$ in the investigated range of $\xi_p$.  
 The mismatch between $\eta_0$ and $\eta_{local}$ indicates that the CNC do not perceive the surrounding medium as a continuum, thus experiencing a viscosity that lies between the water viscosity ($\eta_{H_2O}$) and the macroscopic bulk viscosity $\eta_0$ of the polymer solutions. Consequently, predicting the minimum shear rate required to induce CNC alignment in polymer crowds, based on the criterion $\dot\gamma=Dr$ (i.e., $Pe=1$), using the bulk viscosity $\eta_0$ as the solvent viscosity in eqn.~\ref{eqn:Dr}, would fail by underestimating $Dr$. 
%
\begin{figure}[ht!]
\centering
\includegraphics[width=9cm]{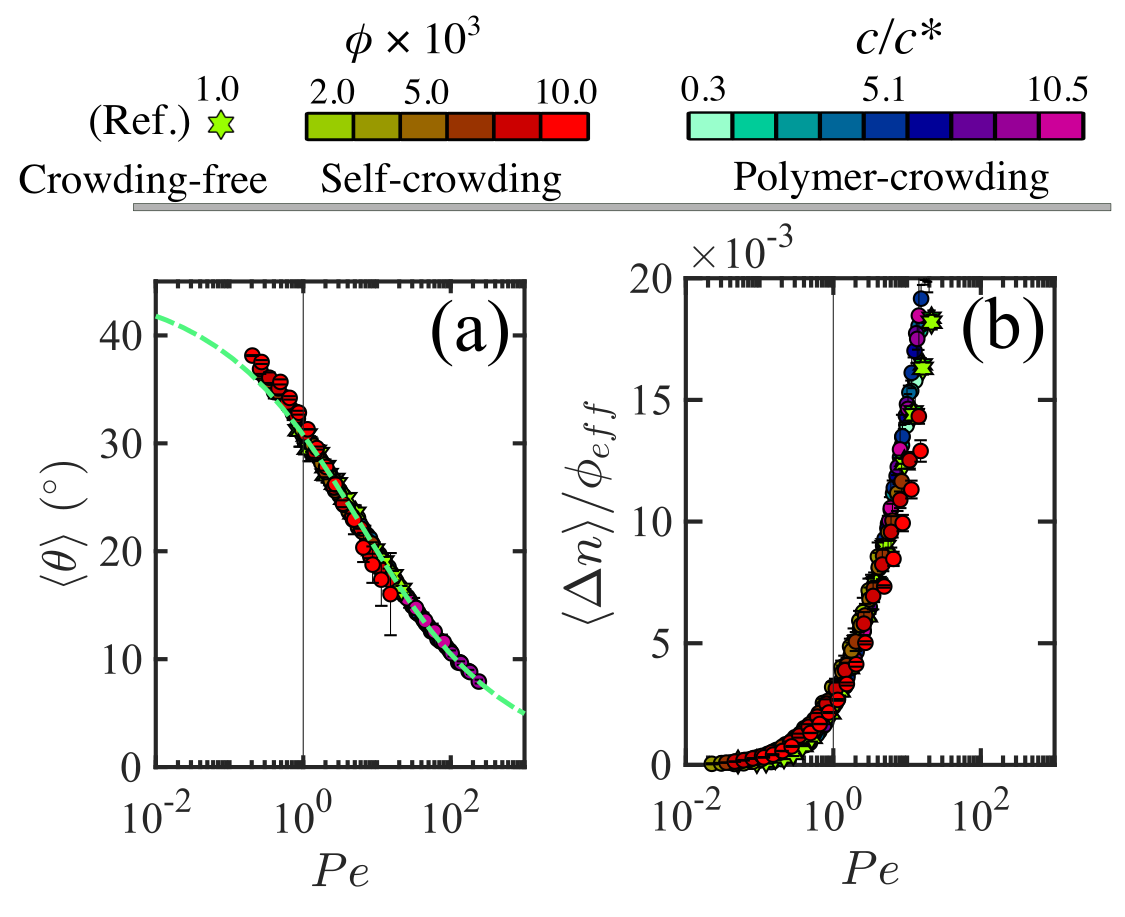}
\caption{(a)  Orientation angle, $\langle \theta \rangle $, and normalized birefringence, $\langle \Delta n \rangle / \phi_{eff}$, as a function of $Pe$ for the self-crowding (from Fig.~\ref{fgr:1}(f,~h)) and polymer-crowding  (from Fig.~\ref{fgr:2}(b,~e) of the main manuscript) for a total of 16 data sets including the reference sample. The dashed line in (a) is the plot of eqn.~\ref{eqn:Orientation Angle} using $\alpha=0.39$.}
\label{fgr:Master Curve}
\end{figure}

\subsubsection{Time-scale dependence}

Flow curves obtained from bulk shear rheometry were used to probe the two crowding agents, CNC suspensions in water and PEO8 solutions in the absence of CNC, under different shear rates and related flow time scales (i.e., $t=\dot\gamma^{-1}$), see Fig.~\ref{fgr:4}(a,~b). It is important to note that the presence of CNC at $\phi=1.0 \times 10^{-3}$ in the PEO8 solutions did not alter the bulk rheology (see ESI, Fig.~S1). With increasing concentrations of the crowding agent ($\phi$ and $c/c^*$ for the CNC and PEO8, respectively), the shear viscosity increases and the onset of shear-thinning, captured by $\dot\gamma^*$ from the CY model (eqn.~\ref{eqn:Carreau_Yasuda}), shifts to lower values of shear rate (depicted by the yellow squares in Fig.~\ref{fgr:4}(a, b)). The $Dr$ obtained via birefringence measurements (as shown in Fig.~\ref{fgr:3}(a, b)) are marked  as triangular red symbols on the respective plots of the CNC suspensions and PEO8 solutions.

For the CNC suspensions (Fig.~\ref{fgr:4}(a)), $\dot\gamma^*$ obtained from shear rheometry and $Dr$ obtained via birefringence measurements have similar values because the onset of shear-thinning correlates with the onset of CNC alignment. Therefore, the physical interpretation for the decreasing $\dot\gamma^*$ as a function of $\phi$ mirrors the one given for $Dr$. Specifically, in the semi-dilute (self-crowding) regime, the rods perceive the surrounding rods, enabling the onset of alignment at values of shear rate that decrease progressively with increasing $\phi$. 
For the PEO8 solutions (Fig.~\ref{fgr:4}(b)), the onset of shear-thinning,  $\dot\gamma^*$, corresponds to the longest relaxation time of the polymer solution as $\tau_p =\dot\gamma^{*-1}$. In good approximation, the CNC alignment in PEO8 occurs at a shear rate $\dot\gamma=\dot\gamma^*=Dr_0$ (see yellow squares and red triangles in Fig.~\ref{fgr:4}(b)), suggesting that the onset of CNC alignment is coupled with the polymer relaxation time as $\tau_p = Dr_0^{-1}$. As such, a suitable control parameter for the CNC alignment is the Weissenberg number, $Wi=\tau_p\dot\gamma$, quantifying the strength of the elastic response of the fluid to an imposed deformation rate, where for $Wi<1$ the polymers are in their equilibrium conformation whilst for $Wi\geq 1$ the relatively high flow rate drives the polymers out of their equilibrium conformation.\citep{Nikoubashman2017,Ebagninin2009,Dunderdale2020} Based on the relationship $\tau_p = Dr_0^{-1}$, the Weissenberg and P\'eclet numbers become identical because the rotational diffusion time-scale of the CNC rods is the same as the longest relaxation time of the polymer for the regime $\xi_p<Rg_r<t_d$ investigated here (analogous results are also obtained for the PEO4 and PA5 solutions presented in ESI, Fig.~S4). Consistently, the trend of CNC alignment in PEO8 solutions captured by $\langle \Delta n \rangle /\phi$ and $\langle \theta \rangle$ as a function of $Pe$ is also well described by $Wi$, with the onset of CNC alignment occurring at $Pe=Wi=1$ (Fig.~\ref{fgr:4}(c,~d)). This is remarkably different from previous reports of elliptical hematite particles suspended in PEO solutions in the regime $Rg_r > t_d$, where the onset of alignment occurs at  $Wi\ll 1$ \cite{Gunes2008}. Similarly, the onset of alignment of carbon nanotubes in sheared polymer melt was observed at $Wi\ll1$, ruling out the coupling of $\tau_p$ with $Dr$ for relatively large colloids ($Rg_r > t_d$).\cite{Hobbie2003}
From a topological perspective, the coupling between tracer particles and the polymer dynamics is predicted by Cai et al.\cite{Cai2011} in the regime $\xi_p<Rg_r<t_d$ with the scaling $Dr \propto \xi_p^2$ (see Fig.~\ref{fgr:3}(b)). It is possible to conceptualize the coupling of $Dr$ with $\tau_p$ by analysing three distinct time-scales at play during flow as conceptualized in Fig.~\ref{fgr:4}(e). For $t > \tau_p$ ($\dot\gamma<\dot\gamma^*$) the polymer is relaxed and in its equilibrium configuration as the probed time scales are long enough to enable polymer relaxation, during which the CNC is able to escape from the transient confinement provided by the polymer mesh, thus Brownian diffusion dominates. Increasing the shear rate we reach $t=\tau_p$($\dot\gamma=\dot\gamma^*$) where the polymer is driven out of its equilibrium conformation and deformed by the flow. At this time scale the CNC perceives the surrounding polymer as a static mesh that provides confinement and aids CNC alignment at values of $\dot\gamma^* = Dr$ ($\tau_p= Dr^{-1}$). At higher values of shear rate, $t<\tau_p$ ($\dot\gamma>\dot\gamma^*$),  the CNC continues to align with the flow, following a universal curve of $\langle\theta \rangle$ and $\langle\Delta n \rangle /\phi_{eff}$ with respect to $Pe$ for a wide range of polymer concentrations in the semidilute unentangled regime as shown by the master curves in Fig.~\ref{fgr:2}(b,~e). 
We note that the CNC alignment for the polymer- and self-crowding case display analogies. In Fig.~\ref{fgr:Master Curve} we plot together $\langle\theta \rangle$ and $\langle\Delta n \rangle /\phi_{eff}$ as obtained for all the polymer- and self-crowding cases presented above. The CNC alignment can be described by the same master curve of $\langle\theta \rangle$ and $\langle\Delta n \rangle /\phi_{eff}$ vs.~$Pe$. This observation implies that both self- and polymer-crowding of the CNC only alters the critical shear rate for the onset of alignment, but the subsequent trend in alignment for $Pe>1$ remains the same for both cases. This universal trend with respect to the crowding agent is likely caused by the inability of the CNC to explore the surrounding confinement once the alignment is triggered by a sufficiently high shear rate (i.e., $Pe=1$).\cite{Lang2019} 
Note that in our present study the characteristic polymer time-scale is greater than the rotational diffusion time-scale of the CNC in the crowding-free regime, $\tau_p > Dr_0^{-1}$ ($\dot\gamma^* < Dr_0$), see vertical line in Fig.~\ref{fgr:4}(b).
However, colloidal rods with slower rotational dynamics in the crowding-free regime compared to the characteristic polymer time-scale, will be in the regime $\tau_p < Dr_0^{-1}$ ($\dot\gamma^* > Dr_0$). Practically, this regime can be achieved by increasing the length of the colloidal rods ($Dr_0 \propto l^{-3}$), and/or decreasing the polymer molecular weight. As the regime $\tau_p < Dr_0^{-1}$ is approached, by for instance increasing $l$, the colloidal rods will progressively experience the surrounding environment as a continuum rather than a discrete medium and $\eta_{local} \rightarrow \eta_0$. Therefore, for $\tau_p \ll Dr_0^{-1}$ ($\dot\gamma^* \gg Dr_0$) the onset of colloidal alignment is expected to be dominated by the bulk viscosity of the surrounding polymer solution.\cite{Wyart2000}


\subsection{Conclusions}  
  
We tackle an industrially relevant problem from a fundamental perspective: the control over the alignment of rod-like colloids in polymeric matrices. Specifically, we compare the flow-induced alignment of rigid colloidal rods, namely CNC, in two contrasting crowded environments referred to as self- and polymer-crowding. By analysis of the length- and time-scale, we find that rotational diffusion coefficient, $Dr$, of CNC in high molecular weight polymeric crowds is coupled with the longest relaxation time of the surrounding polymer, $\tau_p$. On this ground, we propose the Weissenberg number $Wi$ as the control parameter for the alignment of colloidal rods that possess similar length-scales as the suspending polymer fluid, $\xi_p <Rg_r< t_d$; i.e., in conditions where the continuum approach breaks down. Specifically, we show that by knowing $\tau_p$ from rheological tests, it is possible to predict the critical shear rate for the onset of colloidal alignment in polymeric fluids as $\dot\gamma=\tau_p^{-1}$, equivalently $Wi=1$, without relying on the knowledge of the local viscosity experienced by the colloidal rods, $\eta_{local}$. In conclusion,  our results provide crucial insights on the dynamics of colloidal rods under shearing flows that will aid the production of composite materials with desired structural organization. Additionally, the ability of tracer colloidal rods to probe the relaxation times of the surrounding polymers opens the opportunity to perform \textit{in-situ} and spatially-resolved characterisation of the dynamics of polymeric fluids under flow using tracer colloidal rods. With further optimisation (e.g., size and composition of the colloidal rods), this technique is promising for analysing polymer dynamics in complex flows encountered in real-life conditions where the investigation is  a significant challenge. As a natural next step to our work, we envisage future studies where self- and polymer-crowding are at play jointly, mirroring actual industrial conditions. We use CNC as industrially relevant colloidal rods but the basic principles will also apply to other anisotropic, rod-like, colloidal particles. 

\section{Acknowledgements}
The authors gratefully acknowledge the support of Okinawa Institute of Science and Technology Graduate University with subsidy funding from the Cabinet Office, Government of Japan. V.C, SV, S.J.H. and A.Q.S. also acknowledge financial support from the Japanese Society for the Promotion of Science (JSPS, Grant Nos. 22K14738, 22K14184, 18K03958, 18H01135, and 21K03884) and the Joint Research Projects (JRPs) supported by the JSPS and the Swiss National Science Foundation (SNSF).


\section{Competing interests}
The authors declare no competing interests.

\section{Additional information}
Supplementary information is available for this paper.


\providecommand{\latin}[1]{#1}
\makeatletter
\providecommand{\doi}
  {\begingroup\let\do\@makeother\dospecials
  \catcode`\{=1 \catcode`\}=2 \doi@aux}
\providecommand{\doi@aux}[1]{\endgroup\texttt{#1}}
\makeatother
\providecommand*\mcitethebibliography{\thebibliography}
\csname @ifundefined\endcsname{endmcitethebibliography}
  {\let\endmcitethebibliography\endthebibliography}{}

\end{document}